\documentclass{article}
\usepackage{spconf,amsmath,graphicx,hyperref}

\usepackage{cite}
\usepackage{amsmath,amssymb,amsfonts}
\usepackage{algorithmic}
\usepackage{textcomp}
\usepackage{xcolor}
\usepackage{comment}
\usepackage{arydshln}
\usepackage{booktabs} % 表の装飾  
\usepackage{multirow} % セルの結合  
\usepackage{graphicx} % 表のスケーリング用  
\usepackage{tipa}
\usepackage{physics}
\usepackage{subcaption}
\usepackage{caption} 
\usepackage{enumitem}
\usepackage{graphicx} 

\newcommand{\system}[2]{\texttt{#1}-\texttt{#2}}

\newcommand{\customdashline}[1]{%
  \noalign{\vskip\aboverulesep} % 上のスペース
  \cdashline{#1}
  \noalign{\vskip\belowrulesep} % 下のスペース
}

\title{Synthetic Data Domain Adaptation for ASR via LLM-based Text and Phonetic Respelling Augmentation}
\name{Natsuo Yamashita, Koichi Nagatsuka, Hiroaki Kokubo, Kota Dohi, Tuan Vu Ho}
\address{Hitachi, Ltd.}
\begin{document}
\ninept
% \setlength{\parskip}{0pt}
%\ninept
%
\maketitle

\begin{abstract}
End-to-end automatic speech recognition often degrades on domain-specific data due to scarce in-domain resources. We propose a synthetic-data-based domain adaptation framework with two contributions: (1) a large language model (LLM)-based text augmentation pipeline with a filtering strategy that balances lexical diversity, perplexity, and domain-term coverage, and (2) phonetic respelling augmentation (PRA), a novel method that introduces pronunciation variability through LLM-generated orthographic pseudo-spellings. Unlike conventional acoustic-level methods such as SpecAugment, PRA provides phonetic diversity before speech synthesis, enabling synthetic speech to better approximate real-world variability. Experimental results across four domain-specific datasets demonstrate consistent reductions in word error rate, confirming that combining domain-specific lexical coverage with realistic pronunciation variation significantly improves ASR robustness.
% The augmentation pipeline enhances lexical diversity using filtering strategies based on Type-Token Ratio (TTR), perplexity, and domain-specific terms, while the respelling method introduces pronunciation variability during text generation to simulate realistic speech patterns.
% The augmentation pipeline increases lexical diversity and filter domain-relevant text that enhances the lexical and contextual coverage

% The phonetic respelling method introduces pronunciation variability directly at the text generation stage, better simulating realistic speech variations than spectrogram-level augmentation for synthesized audio.

% Generated text and audio samples are available at \url{https://natsuooo.github.io/LPR/}.
\end{abstract}


\section{Introduction}
\label{sec:intro}

End-to-end automatic speech recognition (ASR) systems have achieved remarkable progress in recent years, but they still suffer substantial performance degradation when applied to domain-specific data that differs from the training distribution~\cite{review}.
% TTSベースの手法＋LLMで生成した文章->TTS
% Since collecting large amounts of target-domain speech is often impractical, text-to-speech (TTS)-based synthetic data generation has been explored as a cost-effective alternative for ASR domain adaptation~\cite{das, dysarthric_asr}.
Since collecting large amounts of target-domain text and speech can be costly, recent studies have explored generating domain-specific text using large language models (LLMs) and converting it into synthetic speech via text-to-speech (TTS) as a cost-effective approach for ASR domain adaptation~\cite{das, dysarthric_asr}.
However, the existing synthetic-data approaches face two key limitations:
(1) insufficient domain-specific lexical diversity---these studies have primarily focused on increasing the amount of text without explicitly optimizing for domain-aware lexical diversity and coverage;
and (2) lack of natural phonetic variability---synthetic speech generated via TTS lacks the pronunciation variations, errors, and idiosyncrasies found in real speech~\cite{tts_mono_2}. 
Existing acoustic-level augmentation methods (e.g., SpecAugment~\cite{specaug}) mask parts of the spectrograms rather than introduce pronunciation variants and, when applied to uniformly rendered synthetic speech, can be detrimental in some setups~\cite{dysarthric_asr}.
% While acoustic-level augmentation methods such as SpecAugment~\cite{specaug} perturb the spectrogram but do not alter  pronunciation variants, and they often prove ineffective or even detrimental when applied to uniform synthetic speech~\cite{dysarthric_asr}.
% Additionally, a previous study investigated applying SpecAugment~\cite{specaug} during fine-tuning on synthetic speech, but reported a degradation in recognition accuracy~\cite{dysarthric_asr}.
% This degradation could be due to synthetic speech being overly clean and highly uniform, lacking the variability and diversity found in real speech~\cite{tts_mono}, making time- or frequency-masking distortions less beneficial or even harmful for model learning.

\begin{figure}[t]
    \centering
    % Subfloat 1: First image with caption
    \subfloat[Text augmentation pipeline]{%
        \begin{minipage}[b]{1.0\linewidth}
            \centering
            \includegraphics[width=1\columnwidth]{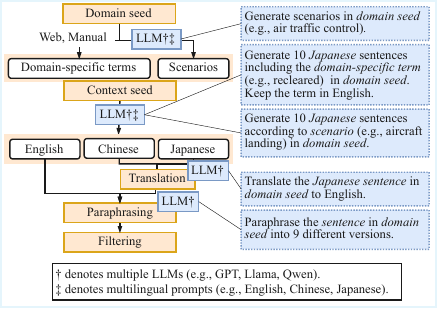}
        \end{minipage}
        \label{fig:overview_1}
    }
    
    % \quad % Adds space between subfigures
    \vspace{0.5em}

    % Subfloat 2: Second image with caption
    \subfloat[Phonetic respelling augmentation]{%
        \begin{minipage}[b]{1.0\linewidth}
            \centering
            \includegraphics[width=1\columnwidth]{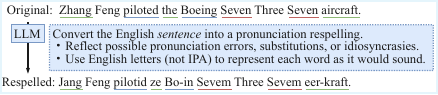}
        \end{minipage}
        \label{fig:overview_2}
    }
    
    % \caption{Overview of our proposed method. Italics indicate placeholders.}
    \caption{Proposed methods overview. Italics denote placeholders.}
    \label{fig:overview}
\end{figure}

% 本研究では，コスト削減のために専門用語リストのみが取得可能であり
% To address these limitations, we propose a robust ASR domain adaptation framework based on synthetic data that (1) enhances domain-specific lexical diversity through an LLM-based text augmentation pipeline with a filtering strategy based on maximizing Type-Token Ratio (TTR) and perplexity, and weighting domain-specific vocabulary, and (2) improves phonetic variability by introducing an LLM-based phonetic respelling method. 
% Unlike SpecAugment, which modifies spectrograms afterwards, our approach injects pronunciation diversity at the text generation stage, simulating realistic variations such as pronunciation errors, substitutions, or idiosyncrasies.
To address these limitations, we propose a robust ASR domain adaptation framework that relies solely on synthetic data. Our approach makes two key contributions:
(1) it enhances domain-specific lexical diversity via an LLM-based text augmentation pipeline, equipped with a novel filtering strategy that jointly maximizes type–token ratio (TTR), perplexity, and domain-specific vocabulary coverage; and (2) it introduces phonetic respelling augmentation (PRA), a novel method that leverages LLMs to generate orthographic pseudo-spellings reflecting realistic pronunciation variability.
Unlike SpecAugment, which modifies acoustic features after synthesis, PRA injects phonetic diversity directly at the text stage, enabling synthetic speech to capture natural variations such as pronunciation errors, substitutions, and idiosyncrasies, while remaining fully compatible with standard TTS systems.
% This yields more natural and varied synthetic speech, improving ASR robustness in low-resource target domains.

% Experimental results show that our LLM-based text augmentation pipeline improves recognition accuracy for various domain-specific datasets, and our phonetic respelling provides further improvements.
% Prompt details, generated texts, and synthesized audio samples are available on our project page.\footnote{\scriptsize\url{https://natsuooo.github.io/llm-asr-augmentation/}} 
Experimental results show that our LLM-based text augmentation pipeline improves recognition accuracy across multiple domain-specific datasets, and PRA yields additional gains.
Example prompts, generated texts, filtering code, and audio samples with multiple TTS systems are available on our project page.\footnote{\scriptsize\url{https://natsuooo.github.io/llm-asr-augmentation/}} 

% 先行研究では，LLMを用いてテキストを生成し，音声合成してファインチューニングすることでドメイン適応するフレームワークを示した。しかし，単に上記を用いるだけでは，多様性に欠ける文章生成となっている。

% ターゲットドメインに関するデータは，入手できない前提。ただし，多くのビジネスケースで，対応したい専門用語リストが与えられるため，rare wordsのみ用いることができるケースを想定する

% DASは，domain seed+context seed -> 本研究では，複数LLM, rewrite, filteringが新しい

\section{Related work}
% \subsection{Increasing textual diversiy with LLMs}
% % Recent studies have explored the use of LLMs to create first meta data and enhance the textual diversity by generating instructions from metadata to downstream ASR domain adaptation~\cite{das, dysarthric_asr}. 
% Recent studies have explored the use of LLMs to enhance textual diversity for ASR domain adaptation, for example by first generating scenario-based seeds and then generating texts for each seed~\cite{das, dysarthric_asr}.
% Multilingual prompting has also proven effective for enriching linguistic and cultural diversity in synthetic datasets~\cite{multilingual}. 
% Furthermore, LLMs have demonstrated the ability to generate paraphrases that not only preserve semantic similarity but also introduce greater diversity in vocabulary and sentence structure~\cite{paraphrasing}. 
% Ensemble methods that combine outputs from multiple LLMs have been found to further boost the diversity and quality of generated content~\cite{multiple_llm}.
% However, most prior work has focused on the diversity of generated text, with less attention paid to the efficient selection of relevant and diverse subsets from generated texts for downstream ASR domain adaptation.

\subsection{Text filtering for ASR domain adaptation}
Recent studies have explored the use of LLM-generated text for downstream ASR domain adaptation~\cite{das, dysarthric_asr} but the quality of the text has not been sufficiently discussed.
Some previous studies have focused on filtering out high-perplexity sentences, aiming to retain only fluent texts~\cite{filtering_ppl_2, filtering_ppl}. 
However, prioritizing low-perplexity sentences may risk excluding specialized or technical expressions necessary for domain adaptation.
Other approaches employ vocabulary coverage maximization (VCM) to ensure lexical diversity by maximizing the number of unique words included in the training set~\cite{ho}.
Nevertheless, VCM may lead to the selection of numerous irrelevant or meaningless words generated by LLMs, which can hinder model learning rather than enhance it.
% To address these limitations, we propose filtering based on maximizing both TTR and perplexity with weighting domain-specific vocabulary.
In contrast, we adopt a normalized tri-objective selection that jointly balances TTR, perplexity, and domain-term coverage, resulting in texts that are both diverse and domain-specific.

\subsection{Phonological Tasks and LLMs}
% Traditional grapheme-to-phoneme (G2P) conversion models typically operate at the word level and rely heavily on pronunciation dictionaries, which leads to struggle with context-dependent pronunciations and out-of-vocabulary words. 
% Recent research has explored employing LLMs for phonological tasks such as G2P conversion~\cite{phonologybench, llm_g2p}, but their performance still lags behind dedicated phonological models or human annotators. 
% As a result, directly using phoneme sequences produced by LLMs as input to TTS systems which can convert phoneme to speech, can lead to poor synthesis quality.
% To overcome this issue, instead of using special symbols like the International Phonetic Alphabet (IPA) or other complex phonetic notations, we propose to represent pronunciation variations by LLM-based phonetic respellings written with ordinary alphabetic characters as Figure 1 (b) shows.

% Traditional Grapheme-to-Phoneme (G2P) models rely on dictionaries and struggle with domain-specific words.
% LLMs have been explored for phonological tasks such as G2P~\cite{phonologybench,llm_g2p}, but their performance still falls short of specialized models or human annotators. 
% Moreover, complex phoneme representations often degrade Text-to-Speech (TTS) quality. 
% To address these issues, we instead represent pronunciation variations with ordinary alphabetic respellings with LLMs rather than special phonetic symbols like the IPA (see Section 3.2).

Traditional grapheme-to-phoneme (G2P) models rely heavily on pronunciation dictionaries and often fail to generalize to domain-specific words~\cite{g2p_survey}. 
Recently, LLMs have been explored for phonological tasks such as G2P conversion~\cite{phonologybench, llm_g2p}. 
While LLMs can capture broader contextual and linguistic cues than conventional models, prior work has shown that their performance remains inferior to specialized G2P systems or human annotators. 
Moreover, approaches that employ complex phoneme representations (e.g., IPA~\cite{ipa}) often degrade text-to-speech (TTS) quality due to the difficulty of accurately rendering fine-grained phonetic symbols~\cite{yamashita}. 
In contrast, our approach uses LLMs to generate alphabetic respellings that reflect actual pronunciation, thus avoiding the limitations of dictionary-based G2P and specialized phonetic symbols while enabling diverse and realistic variations.

\section{Proposed method}
% Figure~1 shows our proposed text augmentation pipeline and phonetic respelling with the main part of prompts to LLMs.
% First, we generate domain-specific and lexically diverse texts with the proposed pipeline (Section~2.1) and then synthesize audio from the texts.
% Then, we increase phonetic diversity with the proposed phonetic respelling method (Section~2.2) and finally finetune an ASR model with the fully synthesized data.
% Figure~1 illustrates our proposed text augmentation pipeline and phonetic respelling approach, along with sample excerpts from the actual prompts used for LLMs.
% First, we generate domain-specific and lexically diverse texts using the proposed pipeline (Section~2.1), and then synthesize audio from these texts. 
% Next, we enhance phonetic diversity with the proposed phonetic respelling method (Section~2.2). 
% Finally, we fine-tune an ASR model using the fully synthesized data.
% Our approach consists of two main components: an LLM-based text augmentation pipeline (Figure~\ref{fig:overview}(\subref{fig:overview_1})) and an LLM-based phonetic respelling (Figure~\ref{fig:overview}(\subref{fig:overview_2})).
% Synthesized speech from these steps is used to fine-tune the ASR model.
Our framework has two key components: an LLM-based text augmentation pipeline (Figure~\ref{fig:overview}(a)) and PRA (Figure~\ref{fig:overview}(b)). Synthesized speech generated from both components is used to fine-tune the ASR model.

\subsection{LLM-based text augmentation pipeline}
Unlike DAS~\cite{das}, which generates only the required amount of text, our proposed pipeline first over-generates a large pool of candidate sentences through multi-stage LLM-based augmentation (steps 1--5), and then applies a novel filtering process (step 6) to select the most relevant and diverse subset.
% This over-generation and subsequent filtering constitute a key novelty of our approach, enabling more effective control over lexical diversity and domain coverage.
This over-generation and filtering is a key novelty, allowing more effective control over lexical diversity and domain coverage.
% The text augmentation pipeline effectively maximizes domain relevance and lexical diversity through the following six steps:
% The text augmentation pipeline effectively maximizes domain relevance and lexical diversity through the following six steps:
% We propose a comprehensive text augmentation pipeline designed to maximize both domain relevance and lexical diversity, while also introducing a novel filtering strategy following six steps (Figure~1~(a)):
% The proposed text augmentation pipeline comprises the following six steps.
% Our novelty is designing the whole pipeline and proposing a filtering method based on TTR, perplexity, and domain-specific terms.
% Our proposed pipeline (Figure~1) generates diverse, domain-specific training data for ASR domain adaptation via the following steps:
% Figure~1 (a) shows our proposed pipeline with the main part of prompts to LLMs.
% Our proposed pipeline (Figure~1) generates diverse, domain-specific training data for ASR domain adaptation via the following steps. After generating text, the selected sentences are synthesized into speech and used for fine-tuning the ASR model.

% \begin{enumerate}[leftmargin=0pt, labelsep=0.3em, itemindent=2em, wide=0pt, itemsep=0em, parsep=0.3em, topsep=0.3em]
% \begin{enumerate}[leftmargin=0pt]
% \begin{enumerate}
\begin{enumerate}[leftmargin=*, align=left]
    
    \item \textit{Domain seed}: 
    % Generate sentences conditioned on domain-specific information to ensure relevance (e.g., air traffic control utterances). 
    % ASRのドメイン適応では，relevanceとdiversityが重要である。
    % このドメインシードによって，あらかじめわかっているドメイン情報を与えることで，生成されるテキストが，そのドメイン関連であることを約束する。
    % 例えば，ATCOSIMであれば，航空管制における管制官による発話など。
    To ensure that the generated sentences are relevant to the target domain, our pipeline begins by conditioning each prompt on domain-specific information, referred to as a domain seed (e.g., air traffic control).
    % To ensure domain relevance, we generate sentences conditioned on domain-specific information (e.g., air traffic control utterances). 
    % This step is for confirming the relevance with the target domain.
    % In ASR domain adaptation, both relevance and diversity are important. By providing a domain seed that supplies known domain information in advance, we ensure that the generated text is relevant to the target domain. 
    % (e.g., utterances in air traffic control).
    % For example, in the case of ATCOSIM, this would include utterances by air traffic controllers in air traffic control scenarios.
    
    \item \textit{Context seed}:
    Scenarios are generated as context seeds, and texts are created for each scenario to increase diversity.
    % In some business cases, it is required to recognize a few domain-specific terms, which can be obtained from sources such as user names, chat logs, websites, or manuals~\cite{yamashita, inter_biasing}.
    In many production settings, stakeholders provide an operational lexicon (e.g., user names, product names, call signs) obtained from chat logs, websites, or manuals, and systems must recognize these terms with high recall~\cite{inter_biasing, yamashita, biasing, biasing_2}.
    Therefore, we also explore using these terms as context seeds to generate texts containing them.

    % Create context seeds based on both representative scenarios in which utterances occur and domain-specific terms, the latter of which can be obtained from sources such as user names, meeting logs, websites, manuals, or user-registered words~\cite{yamashita}. 
    % Scenarios of utterances are generated as context seeds, and texts are created for each scenario to increase diversity. 
    % We first generate scenarios as context seeds and then generate texts for each scenario to increase diversity, as previous studies~\cite{das, dysarthric_asr}. 
    % In many business cases, although collecting target-domain text and speech is costly, it is often necessary to recognize a small number of domain-specific terms. 
    % These terms can be obtained in advance from sources such as user names, meeting chat logs, websites, manuals, or even words registered by users~\cite{yamashita}. 
    % Therefore, in this work, we also explore leveraging domain-specific terms as context seeds and generate texts that include these terms, as illustrated in Figure~1(a).
    % Sentences are then generated for each context seed.
    % https://arxiv.org/pdf/2406.17962v3
    % Create context seeds tailored to different facets of the domain, and subsequently generate sentences for each seed. 
    % We also leverage domain-specific rare terms because a rare terms list can be obtained in advance from sources such as user names, meeting chat logs, websites, manuals, or even words registered by users in many practical cases \cite{}.
    % CodecLM-like？
    
    \item \textit{Multilingual prompting}: 
    For each context, prompts are constructed in multiple languages (e.g., English, Japanese, Chinese), and the outputs are translated back into the target language to enrich linguistic diversity.
    % This approach allows us to incorporate linguistic and cultural expressions from various languages into the synthetic dataset~\cite{multilingual}. 
    To ensure that domain-specific terms are preserved during translation, we add an instruction such as \textit{``Keep the term in English''} to the prompts in the previous step.
    % For each context, prompts are constructed in multiple languages (e.g., English, Japanese, Chinese), translate outputs back to the target language, and enhance linguistic/cultural diversity \cite{}.
    
    \item \textit{Paraphrasing}:
    % LLMs are used to generate paraphrases of existing sentences, thereby increasing the diversity of the training data by providing alternative phrasings for the same content.
    % Following previous findings that LLMs can generate paraphrases which preserve semantic similarity while enhancing lexical and syntactic diversity~\cite{paraphrasing}, we also employ LLM-based paraphrasing in our pipeline
    % Use LLMs to generate paraphrases of existing sentences. 
    % We further increase diversity by employing LLM-based paraphrasing, generating alternative expressions for each generated sentence while preserving semantic content. 
    % LLM-based paraphrasing further increases diversity by generating alternative expressions for each sentence.
    LLM-based paraphrasing generates alternative expressions for each sentence.

    \item \textit{Multiple LLMs}: 
    % 近年のNLPの研究では，複数LLMによって生成した方が多様性が向上したり，それらを集約することでより良質な結果が得られることがわかっている。
    % 以下3, 4, 5のステップは，複数LLMを用いて実行するものとする。
    % The previous three steps (context seed, multilingual prompting, and rewriting) are carried out using multiple LLMs, whose outputs are then combined.
    % Steps 2--4 are carried out using multiple LLMs, whose outputs are aggregated.
    % Steps 2--4 are performed using multiple LLMs, whose outputs are combined to further maximize lexical and syntactic diversity.
    % Steps 2–4 are performed with multiple LLMs and their outputs are combined to maximize lexical and syntactic diversity.
    Multiple LLMs trained on different data are used in Steps 1–4, and their outputs are combined.
    \item \textit{Filtering}: 
    While the above steps yield a large pool of candidate sentences, synthesizing speech and fine-tuning an ASR model on the entire set is inefficient.
    To address this, we introduce a novel filtering step based on three key heuristics: (1) maximizing TTR to promote lexical diversity, (2) maximizing perplexity to encourage technical and in-domain words, and (3) weighting domain-specific terms to enhance coverage of underrepresented vocabulary.
    We compute the combined score $S(s)$ for each candidate sentence $s$ as a weighted sum of TTR gain, perplexity, and the normalized count of domain-specific terms, as follows:
    % \begin{small}
    % \[
    % \alpha\, \frac{|\mathrm{Vocab}(s)\!\setminus V|}{|s|}
    % + \beta\, \exp\left(
    %     -\frac{1}{|s|} \sum_{i=1}^{|s|} \log p(w_i)
    % \right)
    % + \gamma\, \frac{|\{ w_i\!\in s\!\mid\!w_i\!\in \mathcal{D}\!\}|}{|s|}
    % \]
    % \end{small}
    % \begin{align}
    %     S(s) &= 
    %       \alpha\, \frac{|\mathrm{Vocab}(s)\setminus V|}{|s|}
    %       + \beta\, \exp\left(
    %             -\frac{1}{|s|} \sum_{i=1}^{|s|} \log p(w_i)
    %         \right) \notag \\
    %     &\quad + \gamma\, 
    %         \frac{|\{ w_i \in s \mid w_i \in \mathcal{D} \}|}{|s|}.
    % \end{align}
    % \begin{align}
    %     S(s) &=\!\alpha\, \frac{|\mathrm{Vocab}(s)\!\setminus\!V|}{|s|}
    %       + \beta\,\!\exp\left(\!-\frac{1}{|s|} \sum_{i=1}^{|s|} \log p(w_i\!\mid\!w_{<i})
    %         \right) \notag \\
    %     &\quad + \gamma\, 
    %         \frac{|\{ w_i \in s \mid w_i \in \mathcal{D} \}|}{|s|}.
    % \end{align}
    \begin{align}
        S(s) &=\!\alpha\, \frac{|\mathrm{Vocab}(s)\!\setminus\!V|}{|s|}
          + \beta\,\!\exp\left(\!-\frac{1}{|s|} \sum_{i=1}^{|s|} \log p(w_i\!\mid\!w_{<i})
            \right) \notag \\
        &\quad + \gamma\, 
            \frac{|\{ w_i \in s \mid w_i \in \mathcal{D} \}|}{|s|}.
    \end{align}
    where $|\mathrm{Vocab}(s)|$ is the number of unique words in $s$, $V$ is the set of unique words in selected sentences, $|s|$ is the number of words in $s$, $p(w_i)$ is the probability of word $w_i$ (in this context, a token) given the language model, $\mathcal{D}$ is the set of domain-specific terms, and $\alpha, \beta, \gamma$ are weighting coefficients for each term. 
    % Each term in $S(s)$ is min-max normalized across all candidate sentences prior to weighting.
    Each term in $S(s)$ is min–max normalized within the current candidate pool at each greedy step before weighting.
    % Unlike conventional approaches that prioritize low-perplexity sentences, our method deliberately includes higher-perplexity outputs, thereby exposing the ASR model to more domain-specific vocabulary in the target domain.
    % To ensure data quality, we also apply language-specific length constraints and filter out sentences containing invalid or non-target characters as the previous study~\cite{beyond}.
    % We also apply language-specific length constraints and filter out sentences containing invalid or non-target characters to eliminate collapsed output.
    Unlike conventional approaches that filter for low-perplexity to avoid unnatural sentences~\cite{filtering_ppl_2, filtering_ppl}, our method includes higher-perplexity outputs to better cover domain-specific vocabulary.
    Since our texts are generated by high-quality LLMs, fluency and coherence are already ensured.
    To further guarantee quality, we also apply language-specific length constraints and filter out sentences containing invalid or non-target characters, which effectively eliminates collapsed outputs.
    As an example, we observed that the sentence \textit{``Captain Lee informs Denver zone Hotel-Seven for Airbus Two Hundred.''} exhibits a high perplexity ($\approx\!14550$), yet remains fluent in the context of air traffic control.

    % For the selection algorithm, we employ the MUSS (Multilevel Subset Selection) method~\cite{muss}, which efficiently selects a highly relevant and diverse subset by optimizing a weighted combination of TTR, perplexity, and domain-specific terms through a multilevel greedy strategy based on clustering.
    For the selection algorithm, we employ the multilevel subset selection (MUSS) method~\cite{muss}, which efficiently selects a highly relevant and diverse subset through a multilevel greedy strategy based on clustering. 
    Specifically, MUSS first clusters all candidate sentences, then greedily selects a subset of representative sentences within each cluster in parallel. 
    Next, it selects the most important clusters based on aggregated scores, collects all representative sentences from these clusters, and finally performs a global greedy selection on the pooled candidates. 
    Our novelty lies in maximizing $S(s)$ at each selection stage, enabling the resulting subset to be both highly relevant and diverse.

\end{enumerate}

\begin{figure}[t]
  \centering
  \begin{minipage}[t]{1.0\linewidth}
    \vspace{0pt}
    \centering
    \setcounter{table}{-0}
    \captionsetup[table]{skip=4pt}
    \captionof{table}{Summary of evaluation subsets and generated data. Term Ratio is the proportion of test words that are domain-specific terms.}
    \label{tab:eval_data}
    \setlength{\tabcolsep}{2.5pt}
    \resizebox{1.0\linewidth}{!}{
      \begin{tabular}{@{}lrrrrrrrc@{}}
          \toprule
          \begin{tabular}{@{}l@{}}\\Dataset\end{tabular} &
          \begin{tabular}{@{}r@{}}\\Utts.\end{tabular} &
          \begin{tabular}{@{}r@{}}\\Terms\end{tabular} &
          \begin{tabular}{@{}r@{}}Term\\Ratio\end{tabular} &
          \begin{tabular}{@{}r@{}}\\Scenarios\end{tabular} &
          \begin{tabular}{@{}r@{}}Synthetic\\utts.\end{tabular} &
          \begin{tabular}{@{}r@{}}Filtered\\utts.\end{tabular} & 
          \begin{tabular}{@{}r@{}}Filtered\\duration\end{tabular} & \\
          \midrule
          ATCOSIM~\cite{atcosim}      & 1,901 & 44  & 8.4\,\% & 176   & 165K   & 26.7K & 50h \\
          ATCO2~\cite{atco2}          &   871 & 27  & 3.6\,\% & 108   & 121K   & 22.4K & 50h \\
          Court~\cite{court}   & 3,639 & 33  & 0.5\,\% & 132   & 148K   & 21.9K & 50h \\
          MedSyn~\cite{united_medsyn} & 7,906 & 605 & 4.9\,\% & 2,420 & 2900K  & 27.5K & 50h\\
          \bottomrule
      \end{tabular}
    }
  \end{minipage}%
\end{figure}

\begin{table*}[t]
    \centering
    \setcounter{table}{1}
    \captionsetup[table]{skip=4pt}
    \captionof{table}{Results of varied filtering methods across datasets. The best and second best scores are \textbf{bolded} and \underline{underlined}, respectively.}
    \label{tab:results_text}
    \resizebox{1.0\linewidth}{!}{
    \setlength{\tabcolsep}{2.5pt}
    \begin{tabular}{l@{\hspace{10pt}}
        ccccc % ATCOSIM
        ccccc % ATCO2
        ccccc % Court
        ccccc % MedSyn
        }
        \toprule
        & \multicolumn{5}{c}{\vspace{-0.3ex}ATCOSIM}
        & \multicolumn{5}{c}{ATCO2}
        & \multicolumn{5}{c}{Court}
        & \multicolumn{5}{c}{MedSyn} \\
        \cmidrule(l{\tabcolsep}r{\tabcolsep}){2-6}\cmidrule(l{\tabcolsep}r{\tabcolsep}){7-11}\cmidrule(l{\tabcolsep}r{\tabcolsep}){12-16}\cmidrule(l{\tabcolsep}r{\tabcolsep}){17-21}
        % Ours spanning only Pipeline〜PPLmin
        &      & \multicolumn{4}{c}{\vspace{-0.4ex}Ours}
        &      & \multicolumn{4}{c}{Ours}
        &      & \multicolumn{4}{c}{Ours}
        &      & \multicolumn{4}{c}{Ours} \\
        \cmidrule(l{\tabcolsep}r{\tabcolsep}){3-6}\cmidrule(l{\tabcolsep}r{\tabcolsep}){8-11}\cmidrule(l{\tabcolsep}r{\tabcolsep}){13-16}\cmidrule(l{\tabcolsep}r{\tabcolsep}){18-21}
        % Methods row
        & DAS & Pipeline & Rand. & VCM & PPLmin
        & DAS & Pipeline & Rand. & VCM & PPLmin
        & DAS & Pipeline & Rand. & VCM & PPLmin
        & DAS & Pipeline & Rand. & VCM & PPLmin \\
        \midrule
        MATTR
        & 0.784 & \underline{0.884} & 0.875 & \textbf{0.895} & 0.842
        & 0.763 & \underline{0.879} & 0.869 & \textbf{0.882} & 0.844
        & \textbf{0.818} & 0.801 & 0.778 & \underline{0.806} & 0.737
        & 0.902 & \underline{0.916} & 0.902 & \textbf{0.917} & 0.862 \\
        Distinct-2
        & 0.226 & \underline{0.269} & 0.253 & \textbf{0.336} & 0.203
        & 0.220 & \underline{0.246} & 0.238 & \textbf{0.300} & 0.203
        & 0.291 & \underline{0.300} & 0.256 & \textbf{0.328} & 0.171
        & 0.340 & \underline{0.416} & 0.384 & \textbf{0.534} & 0.187 \\
        Perplexity
        & \underline{819.6} & \textbf{1168.0} & 614.8 & 610.5 & 117.7
        & \underline{822.2} & \textbf{970.0} & 464.4 & 464.8 & 121.9
        & \underline{137.1} & \textbf{198.8} & 106.5 & 124.0 & 33.0
        & 118.4 & \textbf{507.3} & 122.9 & \underline{165.1} & 14.6 \\
        Avg.\,Term
        & 0.3 & \textbf{248.5} & \underline{111.2} & 110.5 & 72.2
        & 5.2 & \textbf{345.4} & \underline{136.2} & 120.2 & 100.5
        & 13.4 & \textbf{280.2} & \underline{124.5} & 82.8 & 66.9
        & 1.6 & \textbf{20.4} & \underline{8.3} & 3.9 & 6.2 \\
        \bottomrule
    \end{tabular}
    }
\end{table*}

\subsection{Phonetic respelling augmentation}
As discussed in Section~\ref{sec:intro}, applying SpecAugment to overly clean synthesized speech during ASR fine-tuning can degrade recognition performance.
PRA addresses this by injecting pronunciation variability directly at the text level rather than the spectrogram level.
% Phonetic respelling refers to rewriting words using alternative spellings in the same orthography to approximate natural pronunciation, without resorting to IPA or other specialized phonetic symbols~\cite{respelling}.
PRA rewrites words using alternative spellings in the same orthography to approximate natural pronunciation, without relying on IPA or other specialized phonetic symbols.
Our method generates respellings that capture common spoken phenomena such as assimilation (\emph{Sevem} for \emph{Seven}), elision (\emph{pilotid} for \emph{piloted}, \emph{Bo-in} for \emph{Boeing}), and substitutions (\emph{Jang} for \emph{Zhang}, \emph{ze} for \emph{the}, \emph{eer-kraft} for \emph{aircraft}), using prompts as shown in Figure~\ref{fig:overview}(b).
Note that during training, respelled text is used only as TTS input, while the ASR target remains the original, canonical text.
This approach enables the creation of training text that better reflects natural speaker diversity and spontaneous speech, bridging the gap between canonical written text and real-world pronunciation.
A key advantage of our approach is that these respellings remain in standard alphabetic form and can be directly used by off-the-shelf TTS systems, without the need for phoneme dictionaries or parsers.

\section{Experiments}
\subsection{Experimental setup}

We evaluated our domain adaptation approach on four domain-specific English datasets: ATCOSIM~\cite{atcosim}, ATCO2~\cite{atco2} (air traffic control), and Court~\cite{court} (Indian Supreme Court proceedings), all with human-recorded audio, and MedSyn~\cite{united_medsyn} (pharmaceutical descriptions), which consists of TTS-synthesized audio.
A summary of each evaluation subset is provided in Table~\ref{tab:eval_data}.
For MedSyn, due to the extremely large dataset size, we randomly sampled one-tenth of the data for evaluation.
% Following the general approach of prior studies on rare word evaluation in ASR~\cite{contextualized,rare_words,biasing,yamashita}, we define domain-specific terms as words that do not appear in standard ASR training corpora, namely LibriSpeech~\cite{librispeech}, Common Voice~\cite{commonvoice}, and GigaSpeech~\cite{gigaspeech}.
Consistent with the general practice in prior studies on rare word evaluation~\cite{contextualized,rare_words,yamashita}, we define domain-specific terms as words that do not appear in standard ASR training corpora, namely LibriSpeech~\cite{librispeech}, Common Voice~\cite{commonvoice}, and GigaSpeech~\cite{gigaspeech}.
To mitigate annotation noise, each term was required to appear at least twice in the evaluation set.
For text generation, we used multiple LLMs: {\texttt{GPT-4.1-mini}}\\~\cite{chatgpt}, {\texttt{Llama-4-Maverick-17B-128E-Instruct}}~\cite{llama}, and {\texttt{Qwen3-32B}}~\cite{qwen3}, with temperature/top-p settings of 1.0/1.0 for GPT and Llama, and 0.7/0.8 for Qwen, respectively.
Scenario seeds were generated at four times the number of domain-specific terms.
For each context seed and paraphrasing step, we produced 10 sentences.
Multilingual prompting was performed in English, Japanese, and Chinese.
To prevent hallucination, generated text lengths were limited to 5–200 words for English and 5–100 words for Japanese and Chinese.
In the filtering step, we first obtained sentence embeddings using {\texttt{Qwen3-Embedding-8B}}~\cite{qwen3} and grouped all candidate sentences into 1,000 clusters via k-means clustering.
From each cluster, we selected 200 representative samples, then prioritized clusters with higher aggregated quality scores and sequentially collected samples until the total number of selected sentences reached 60,000.
% From each cluster, we selected 200 representative samples to ensure diversity.
% We then prioritized clusters with higher aggregated quality scores and sequentially collected samples until the total number of selected sentences reached 60,000.
Finally, a global selection was performed to ensure that the total duration of the synthesized speech was 50 hours for each dataset.
Perplexity was computed using GPT-2~\cite{gpt2}.
As a baseline, DAS~\cite{das} was implemented by generating 1,000 context seed scenarios, each with 50 texts using GPT, and using only as many texts as needed to synthesize 50 hours of speech.
% Additionally, for baseline filtering, we used the same algorithm as proposed method, but replaced the filtering step with either random selection, VCM or perplexity minimization (PPLmin).
To compare filtering methods, we conducted experiments by replacing the filtering step in our proposed pipeline with random selection, VCM, or perplexity minimization, while keeping all other procedures the same.
Synthetic speech was generated using kokoro-TTS~\cite{kokoro}, with speaker diversity ensured by randomly selecting from 19 American English speakers for each utterance. 
PRA was performed using GPT.
% The filtering weights ($\alpha$:$\beta$:$\gamma=6\!:\!3\!:\!1$), data duration (50 hours), and respelled data mixture ratio (60\%) were set according to the optimal values obtained from the analyses in Figure~2 and Figure~3.
% \subsection{Experimental Settings}
Throughout our experiments, the filtering weights ($\alpha\!:\!\beta\!:\!\gamma\!=\!6\!:\!3\!:\!1$), data duration (50 hours), and respelled data mixture ratio (60\%) were set according to the optimal values obtained from the analyses in Section~\ref{sec:ablation}.

We fine-tuned {\texttt{Whisper-large-v3-turbo}}~\cite{whisper} as the ASR system using a synthetic dataset, which was split into training and validation sets in a 9:1 ratio. 
Following the results of a previous study~\cite{das}, the encoder was kept frozen during training, as this strategy yielded higher accuracy than updating both the encoder and decoder.
The model was trained for 10 epochs with a batch size of 32 and a learning rate of 1e-5, using the AdamW optimizer~\cite{adamw} with 50 warm-up steps. 
% For SpecAugment, we used two configurations: the standard setting (time masking: 100 frames, frequency masking: 27 bins) and a modest setting (time masking: 30 frames, frequency masking: 13 bins).
For SpecAugment, we used standard (100/27) and modest (30/13) masking configurations for time/frequency.
Model checkpoints were saved at each epoch, and the best model was selected based on word error rate (WER) on the validation set. 
All experiments were conducted on a single NVIDIA H200 GPU.

For evaluating the quality of the generated and filtered texts, we employed four complementary metrics:
MATTR~\cite{mattr} (higher values indicate greater lexical diversity), Distinct-n (higher values indicate greater n-gram diversity), perplexity (higher values indicate greater sentence complexity), and the average frequency of domain-specific terms in the dataset (denoted as Avg.\,Term in Table~2).
ASR performance was assessed using standard WER, B-WER (biased WER for words in the predefined list of domain-specific terms), and U-WER (unbiased WER for words outside the list)~\cite{contextualized}.

% In evaluating the quality of the generated and filtered texts, we employed MTLD and MATTR as metrics for lexical diversity, Distinct-2 for n-gram diversity, perplexity for sentence complexity, and self-BLEU for semantic diversity.
% For ASR performance evaluation, in addition to the standard WER, we report U-WER (unbiased WER measured on words not in the biasing list) and B-WER (biased WER measured on words in the biasing list) \cite{}.

\begin{figure*}[t]
  \centering
  \begin{minipage}[t]{0.64\linewidth}
    % 表を左寄せにする
    \raggedright
    \vspace{0pt} % 上端を揃えるため
    \setcounter{table}{2}
    \captionsetup[table]{skip=4pt}
    \captionof{table}{ASR results for text augmentation methods (WER\,/\,B-WER\,/\,U-WER). ``\texttt{P}'' indicates proposed methods; ``\texttt{B}'' indicates baselines.}
    \label{tab:results_proposal_1}
    \setlength{\tabcolsep}{3pt}
    \centering
    \resizebox{1.0\linewidth}{!}{
      \begin{tabular}{@{}llcccc}
        \toprule
        ID & Method & ATCOSIM & ATCO2 & Court & MedSyn \\
        \midrule
        % \multicolumn{6}{@{}l}{\textbf{Original ASR model}}\\
        \texttt{B0} & Whisper-large-v3-turbo & 28.9\,/\,84.0\,/\,29.5 & 57.1\,/\,73.2\,/\,57.7 & 20.3\,/\,85.6\,/\,20.3 & 10.5\,/\,74.8\,/\,10.6 \\
        % \midrule
        \customdashline{1-6}
        % \multicolumn{6}{@{}l}{\textbf{Fine-tuning with synthetic data}}\\
        \texttt{B1} & DAS~\cite{das} & 28.8\,/\,82.5\,/\,29.4 & 54.2\,/\,73.2\,/\,54.7 & 20.0\,/\,72.8\,/\,20.0 & 9.8\,/\,65.4\,/\,9.8 \\
        \customdashline{1-6}
        \system{P1}{1} & Ours (pipeline) & \textbf{23.9}\,/\,\textbf{40.5}\,/\,\textbf{24.5} & \textbf{47.1}\,/\,\textbf{45.1}\,/\,\textbf{47.6} & \textbf{17.8}\,/\,\textbf{36.8}\,/\,\textbf{17.9} & \textbf{8.8}\,/\,\textbf{32.4}\,/\,\textbf{8.8} \\
        \system{P1}{2} & $\hookrightarrow$ filtering: Random & 26.2\,/\,47.8\,/\,26.8 & 54.6\,/\,51.4\,/\,55.3 & 18.0\,/\,43.9\,/\,18.0 & 9.0\,/\,36.0\,/\,9.0 \\
        \system{P1}{3} & $\hookrightarrow$ filtering: VCM & 24.4\,/\,47.2\,/\,25.0 & 53.8\,/\,52.8\,/\,54.5 & 19.1\,/\,45.6\,/\,19.1 & 9.0\,/\,47.5\,/\,9.0 \\
        \system{P1}{4} & $\hookrightarrow$ filtering: PPLmin & 26.6\,/\,45.7\,/\,27.3& 56.1\,/\,52.1\,/\,56.7 & 25.3\,/\,47.4\,/\,25.3 & 9.7\,/\,57.8\,/\,9.7 \\
        \bottomrule
      \end{tabular}
    }
  \end{minipage}%
  \hfill
  \begin{minipage}[t]{0.35\linewidth}
    \vspace{3pt}
    \centering
    % \vspace*{\fill}
    % \vspace{0.5em}
    % \includegraphics[width=1.0\linewidth]{imgs/wer_cropped.pdf}
    % \caption{WER of proposed pipeline with varied weight ratio.}
    \includegraphics[width=1.0\linewidth]{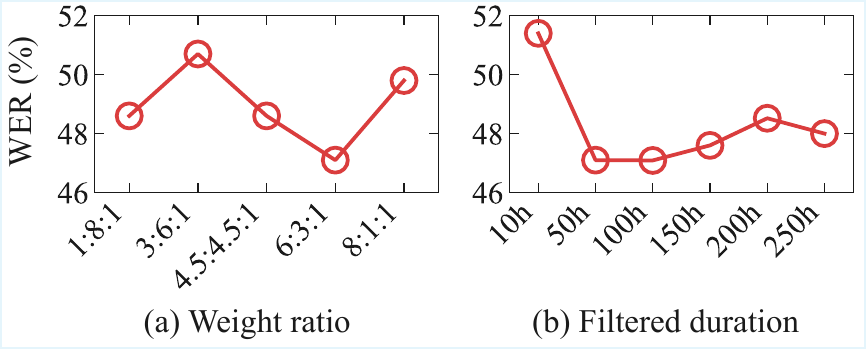}
    \caption{WER with varied weight ratio and filtered duration in \system{P1}{1} on ATCO2.}
    % \vspace*{\fill}
    \label{fig:ablation_1}
  \end{minipage}
\end{figure*}

\begin{figure*}[t]
  \centering
  \begin{minipage}[t]{0.49\linewidth}
    % 表を左寄せにする
    \raggedright
    \vspace{0pt} % 上端を揃えるため
    \captionsetup[table]{skip=4pt}
    \captionof{table}{WER of \system{P1}{1} with SpecAugment and PRA.}
    \label{tab:results_proposal_2}
    \setlength{\tabcolsep}{5pt}
    \centering
    \resizebox{1.0\linewidth}{!}{
      \begin{tabular}{@{}llcccc}
        \toprule
        ID & Method & ATCOSIM & ATCO2 & Court & MedSyn \\
        \midrule
        
        \system{P1}{1} & Ours (pipeline) & 23.9 & 47.1 & 17.8 & 8.8\\
        \customdashline{1-6}
        % \midrule
        % \multicolumn{6}{@{}l}{\textbf{\system{P1}{1}\,+\,Speech augmentation}}\\
        \system{B2}{1} & +\,SpecAugment~\cite{specaug} & 24.3& 44.3 & 21.1 & \textbf{8.6} \\
        \system{B2}{2} & +\,SpecAugment\_modest & 23.4 & 45.4 & 18.7 & 8.8 \\
        \customdashline{1-6}
        \texttt{P2} & +\,Ours (PRA) & \textbf{21.2} & \textbf{41.1} & \textbf{16.8} & 8.7 \\
        \bottomrule
      \end{tabular}
    }
  \end{minipage}%
  \hfill
  \begin{minipage}[t]{0.49\linewidth}
    \vspace{0pt}
    \centering
    % \vspace{0.5em}
    % \includegraphics[width=1.0\linewidth]{imgs/wer_cropped.pdf}
    % \caption{WER of proposed pipeline with varied weight ratio.}
    \includegraphics[width=0.96\linewidth]{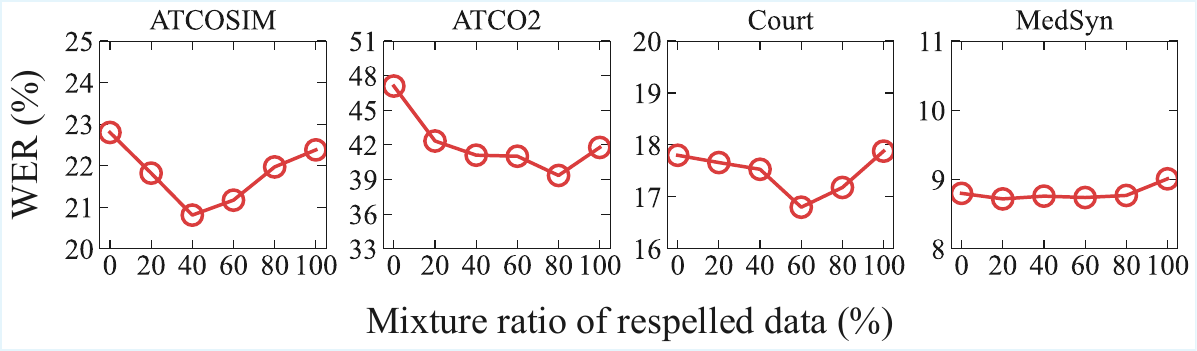}
    \caption{WER with varied mixture ratio of respelled data in \texttt{P2}.}
    \label{fig:ablation_2}
  \end{minipage}
\end{figure*}

\subsection{Preliminary results of text augmentation pipeline}

Table~\ref{tab:results_text} presents the quality of texts using DAS\cite{das} and our pipeline with varied filtering methods: random selection (Rand.), VCM, and perplexity minimization (PPLmin) with MUSS across datasets.
Random selection generally yielded intermediate performance, outperforming DAS in terms of lexical diversity metrics such as MATTR and Distinct-2, but was inferior to VCM and our pipeline.
This improvement over DAS indicates that the 1--5 generation steps are effective in enhancing lexical diversity.
VCM achieved the highest or second-highest scores for metrics reflecting lexical diversity, indicating its strong ability to maximize vocabulary diversity, but it resulted in low perplexity.
In contrast, our pipeline demonstrated high lexical diversity, often ranking just below or comparable to VCM, and achieved the highest perplexity across all datasets.
Perplexity minimization filtering, as expected, showed the lowest perplexity but had the lowest diversity-related scores.
The DAS method generally resulted in lower lexical diversity among all methods.
Regarding average term frequency, our pipeline yielded relatively high values across datasets.
Overall, our pipeline provided a good balance between lexical diversity and perplexity.

\subsection{Main results of ASR performance}
Table~\ref{tab:results_proposal_1} summarizes the ASR performance across four datasets using various text augmentation methods. 
The original Whisper model (\texttt{B0}) shows consistently high WER and B-WER scores, particularly for B-WER, indicating suboptimal performance for domain-specific tasks. 
Fine-tuning with DAS (\texttt{B1}) slightly improves WER, but B-WER improvements remain limited, consistent with previous findings~\cite{das}.
This may be due to low lexical diversity and low perplexity in the training data. 
% In contrast, our filtered pipeline (\texttt{P1-1}) yields consistent reductions in WER, B-WER, and U-WER across all domains.
In contrast, our filtered pipeline (\texttt{P1-1}) yields consistent reductions in WER, B-WER, and U-WER across all domains. 
For \textit{Court}, WER improves from 20.3 to 17.8, B-WER from 85.6 to 36.8, and U-WER from 20.3 to 17.9, despite the domain-term ratio being only 0.5\% (Table~\ref{tab:eval_data}).
% The marked reduction in B-WER demonstrates significant gains in biased word recognition, highlighting the importance of adding domain-specific words as context seeds. 
% The marked reduction in B-WER highlights the importance of both high perplexity and the addition of domain-specific term seeds in improving biased word recognition.
Improvements in WER and U-WER show benefits beyond domain terms, while the large B-WER drop supports combining higher-perplexity selection with domain-term seeding.
VCM (\system{P1}{3}) yields smaller improvements in WER and B-WER compared to the proposed pipeline, suggesting that excessive lexical diversity does not necessarily enhance performance and that high perplexity is also important.
By contrast, perplexity minimization (\system{P1}{4}) sometimes degrades B-WER and U-WER, indicating that selecting only easy texts may hinder domain adaptation.

Table~\ref{tab:results_proposal_2} compares SpecAugment (standard and modest) and PRA (\texttt{P2}) as speech augmentation methods applied to \system{P1}{1}. 
% SpecAugment did not consistently improve WER over the proposed pipeline (\system{P1}{1}) for ATCOSIM, ATCO2, or Court, and sometimes even increased WER. 
% For MedSyn (synthetic speech), SpecAugment achieved the lowest WER (8.6), likely due to added variability in otherwise uniform data, but showed limited benefit for real-speech datasets. 
% In contrast, PRA (\texttt{P2}) consistently achieved the lowest or comparable WERs across all datasets, demonstrating robust domain adaptation by increasing phonetic diversity in the training data.
% SpecAugment did not consistently improve over the pipeline baseline (\system{P1}{1}) on ATCOSIM, ATCO2, or Court and sometimes increased WER. It helped only on MedSyn (synthetic speech). 
% SpecAugment showed effectiveness on the MedSyn dataset, likely because the evaluation data itself also consisted of synthetic speech.
SpecAugment showed effectiveness on the MedSyn dataset, but this is likely because the evaluation data itself also consisted of synthetic speech.
PRA achieved the lowest or comparable WERs across all datasets, indicating that text-level phonetic variability is a reliable augmentation signal for domain adaptation when training on synthetic data.

\subsection{Ablation Study}
\label{sec:ablation}
We analyzed the performance of the proposed pipeline (\system{P1}{1}) by varying the weight ratios ($\alpha\!:\!\beta\!:\!\gamma$) used in filtering and the total duration of filtered synthesized data on ATCO2. 
Figure~\ref{fig:ablation_1}(a) illustrates that the 6:3:1 ratio yielded the best WER (47.1), highlighting the importance of balancing both TTR and perplexity in the filtering process. 
Figure~\ref{fig:ablation_1}(b) presents the results for different amounts of filtered synthesized data. 
When only 10 hours of data were used, WER was higher (51.4). 
However, increasing the data to 50 hours led to a sharp improvement (WER 47.1), and performance remained stable up to 100 hours (WER 47.1). 
Beyond 150 hours, WER gradually worsened, reaching 48.5 at 200 hours and 48.0 at 250 hours.
These findings indicate that while the filtering method effectively selects valuable training data, using excessively large amounts of fully
synthesized data does not consistently improve performance.

% Figure~3 provides further analysis of the proposed phonetic respelling and SpecAugment with the modest setting. 
% The results indicate that a mixture ratio of 40–60\% achieves stable and accurate performance. 
% This is attributed to the fact that the proposed respelling method alters most of the original sentences as an example shown in Figure~1(b), and excessive modifications can hinder training on clean data, emphasizing the importance of a balanced mixing strategy. 
% Additionally, performance degraded when the P2 was combined with SpecAugment, suggesting potential incompatibilities between the two approaches. SpecAugment is most effective when the training and evaluation data exhibit substantial acoustic variability, such as environmental noise or channel effects.
% In contrast, phonetic respelling is particularly useful for synthetic speech datasets, which are typically acoustically uniform but lack pronunciation diversity.

Figure~\ref{fig:ablation_2} provides analysis of the mixture ratio of original and respelled data in PRA (\texttt{P2}).
The results indicate that a mixture ratio of 40\%–60\% achieves stable and accurate performance. 
This is attributed to the fact that PRA alters most of the original sentences, as an example shown in Figure~\ref{fig:overview}(b), and excessive modifications can hinder training on clean data, emphasizing the importance of a balanced mixing strategy.

\section{conclusion}
We present a synthetic-only domain adaptation framework combining LLM-based text augmentation and PRA.
Across four domain-specific datasets, our approach consistently improves B-WER as well as WER and U-WER, showing that injecting lexical and phonetic diversity at the text stage substantially improves ASR robustness for both general and domain-specific vocabulary.
While our experiments focus on English with a single TTS/ASR configuration, our approach is applicable to other languages and speech systems.
We release prompts, code, generated texts, and PRA audio synthesized with multiple TTS systems on our project page, as qualitative references.
Future directions include evaluation with real-world noisy and multilingual data, as well as broader downstream tasks.

% \vfill\pagebreak
\clearpage

\bibliographystyle{IEEEbib}
\bibliography{mybib}

\end{document}